\documentclass[10pt,journal]{IEEEtranTCOM}

\usepackage[english]{babel}
\usepackage[usenames]{color}
\usepackage[cp1250]{inputenc}
\usepackage{amsfonts}
\usepackage{amsthm}
\usepackage{graphicx}
\usepackage{epsfig}
\usepackage{mathrsfs}
\usepackage{amsmath}
\usepackage{algorithm}
\usepackage{algorithmic}
\usepackage{hyperref}
\usepackage{xcolor}
\usepackage{soul}

\theoremstyle{plain}

\begin{document}
\newcommand{\bea}{\begin{eqnarray}}
\newcommand{\eea}{\end{eqnarray}}
\newcommand{\be}{\begin{equation}}
\newcommand{\ee}{\end{equation}}
\newcommand{\beas}{\begin{eqnarray*}}
\newcommand{\eeas}{\end{eqnarray*}}
\newcommand{\bs}{\backslash}
\newcommand{\bc}{\begin{center}}
\newcommand{\ec}{\end{center}}

\def\SC {\mathscr{C}}

\title{Low cost prediction of probability distributions of molecular properties for early virtual screening}
\author{
    \IEEEauthorblockN{Jaros\l{}aw Duda\IEEEauthorrefmark{1}, Sabina Podlewska\IEEEauthorrefmark{2}}\\
    \IEEEauthorblockA{\IEEEauthorrefmark{1}Faculty of Mathematics and Computer Science, Jagiellonian University, Krak\'ow, Poland, \emph{jaroslaw.duda@uj.edu.pl}}\\
    \IEEEauthorblockA{\IEEEauthorrefmark{2}Maj Institute of Pharmacology, Polish Academy of Sciences, Smetna 12, Krak\'ow, 31-343, Poland}}
\maketitle

\begin{abstract}
While there is a general focus on predictions of values, mathematically more appropriate is prediction of probability distributions: with additional possibilities like prediction of uncertainty, higher moments and quantiles. For the purpose of the computer-aided drug design field, this article applies Hierarchical Correlation Reconstruction approach, previously applied in the analysis of demographic, financial and astronomical data. Instead of a single linear regression to predict values, it uses multiple linear regressions to independently predict multiple moments, finally combining them into predicted probability distribution, here of several ADMET properties based on substructural fingerprint developed by Klekota\&Roth. Discussed application example is inexpensive selection of a percentage of molecules with properties nearly certain to be in a predicted or chosen range during virtual screening. Such an approach can facilitate the interpretation of the results as the predictions characterized by high rate of uncertainty are automatically detected.
In addition, for each of the investigated predictive problems, we detected crucial structural features, which should be carefully considered when optimizing compounds towards particular property. The whole methodology developed in the study constitutes therefore a great support for medicinal chemists, as it enable fast rejection of compounds with the lowest potential of desired physicochemical/ADMET characteristic and guides the compound optimization process.
\end{abstract}

\section{Introduction}
\begin{figure}[t!]
    \centering
        \includegraphics[width=85mm]{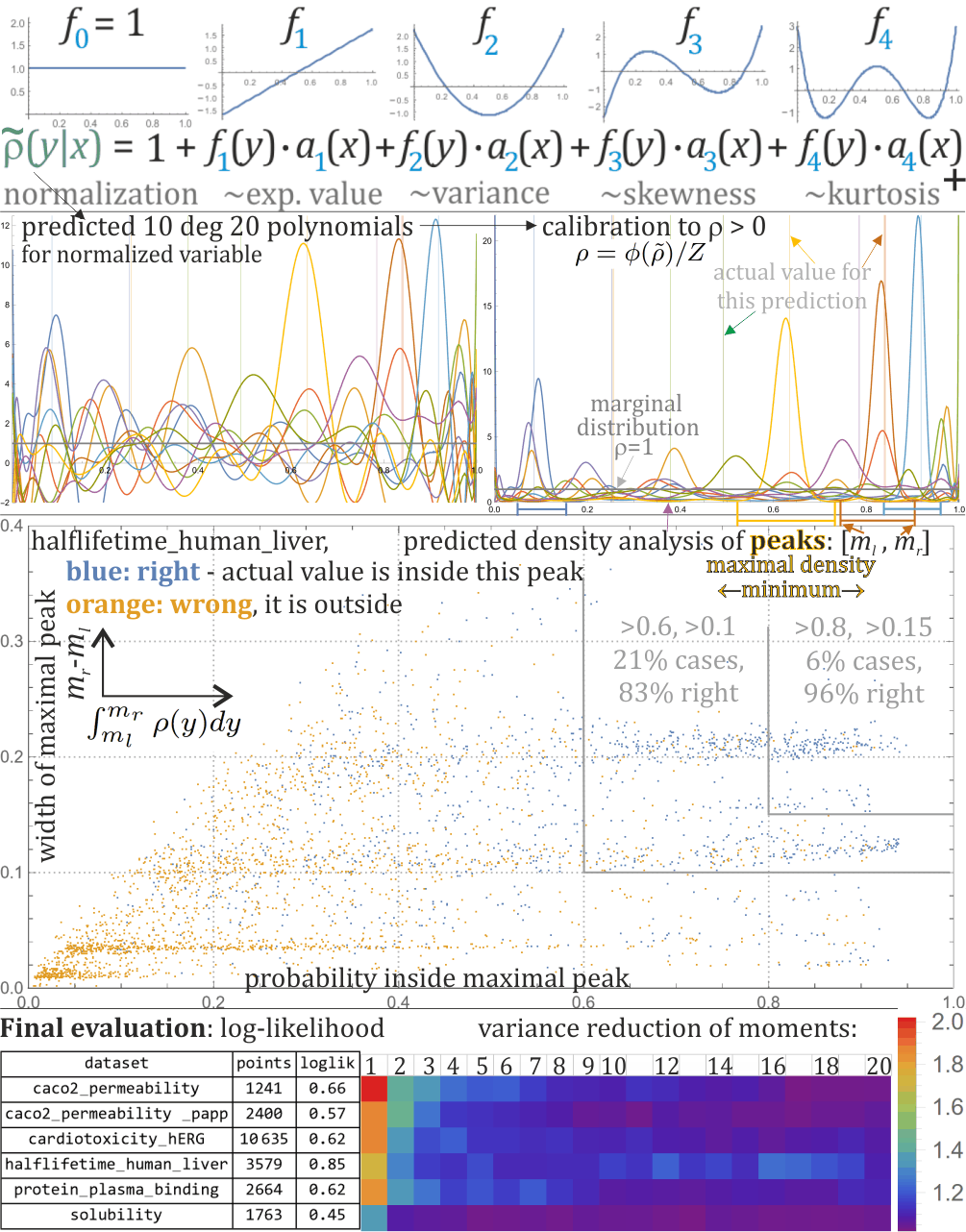}
        \caption{Top: the used HCR-based approach: conditional density for predicted normalized (to nearly uniform in $[0,1]$) variable $Y$ based on features (fingerprints) $X$ is modeled as a linear combination in $(f_i)$ orthonormal basis: $\tilde{\rho}(Y=y|X=x)=1+\sum_{i=1}^m f_i(y) a_i(x)$. The $a_i$ parameters resemble expected value, variance, skewness, kurtosis and further moments - here up to $m=20$, predicted using $m$ independent linear regression of Klek fingerpreints. Density has to be non-negative, ensured by further calibration: $\rho=\phi(\tilde{\rho})/Z$ for $Z$ normalization to integrate to 1, and here $\phi(\rho)=\ln(1+\exp(\mu x)/\nu)/\mu $ family with optimized parameters $\mu,\nu$. Center: as such predicted densities are often narrow peaks, for virtual screening we could focus on them: taking maximum of predicted $\rho$ density and going left/right as long as $\rho$ decreases - getting $[m_l,m_r]$ peak range, we can calculate its width $m_r-m_l$ and probability it contains $\int_{m_l}^{m_r} \rho(y) dy$. Presented scatter plot shows such pairs (probability inside peak, its width) for halflifetime\_human\_liver, visualizing with color if this peak contained the actual value - we can see that restricting to peaks above some thresholds, with very high probability they contain the actual value. Bottom: evaluation of log-likelihoods in 10-fold cross-validation for 6 tested datasets. There is also shown strength of predictions for individual $(a_i)_{i=1..20}$ as original variance divided by variance if subtracting prediction (in 10-fold cross-validation) - we can see that only the first moment ($\sim$expected value) is well predicted, however, there is usually also some prediction for the higher ones.    }
       \label{intr}
\end{figure}
Wide range of computational methods developed to support the process of search for new active compounds contribute to the significant shortening of time and lowering the expenses of the new drugs development pipelines.~(\cite{drug1,drug2}) The intense progress in the computer-aided drug design field is nowadays possible thanks to the continuous gain of knowledge on the human biology, as well as the increase in the available computational power.

The widespread manipulation of chemical data by computers generates the need of the proper representation of compound structures. The most popular way of depicting chemical structure is its representation in the form of molecular descriptors or fingerprints. The former approach, calculates values of various compound parameters, such as the number of atoms of particular type, molecular weight, number of bonds of the given order, etc. On the other hand, fingerprints inform about the presence of particular chemical moieties in the molecule. There are two main fingerprint types: hashed and key-based ones. During the calculation of the hashed fingerprint, each atom constitutes a starting point of a path of a particular length. Then, for each string, a hash code is generated, which finally ends up with the string representing the whole molecule. On the other hand, in the key-based fingerprints, each position has its interpretation, as it represents the presence or absence of the given feature (being most often chemical fragments) in the molecule.

In this article we discuss the application of key-based fingerprints to predict not only the expected value, but also independently moments resembling variance, skewness, kurtosis and higher, finally combining them into a single predicted conditional probability distribution with Hierarchical Correlation Reconstruction (HCR) approach for selected physicochemical and ADMET compound properties. The above-mentioned methodology was previously used for various applications e.g. demographic~\cite{hcr1}, financial~\cite{hcr2}, astronomical~\cite{hcr3}. Here we often have more features than datapoints, requiring new regularization techniques proposed in this article.

Taking into account the information provided by the presented method, the potential application of the predicted probability distribution in computer-aided drug design tasks (virtual screening in particular) is discussed. Combinatorial space of chemical molecules grows exponentially with size, requiring computationally inexpensive methods to select a small percentage of promising ones during search in this space - for example of some properties to be in a desired range, allowed by  proposed quantile thresholding and maximal peak analysis.

In the study, we focus on the prediction of selected physicochemical and ADMET compound features: intestinal permeability (examined via two Caco-2 permeability assays with "papp" referring to the apparent permeability), cardiotoxicity (described as the hERG channel blockage), metabolic stability (investigated as compound half-lifetime in human liver cells), plasma protein binding and solubility. The importance of their (as well as other physicochemical/ADMET properties) is undeniable. Focusing only on compound activity towards considered targets is insufficient, as even the most active compound cannot become a drug, if its physicochemical and ADMET profile is unfavorable. For example, if the metabolic stability of a compound is too low, it has not sufficient time to trigger the desired biological response. Moreover, the decomposition products formed can not only be inactive towards the target, but they can also display toxic effects. Intestinal permeability is important for orally administered drugs to effectively get to the place of action and toxicity is an obvious feature which disqualifies a compound from being a future drug.

Despite the constant increase in the amount new experimental data, which can be used to the development of new predictive models, there still a lot of difficulties, which cause that the predictions are not always sufficiently accurate.

In the study, we propose the non-standard way of dealing with such prediction uncertainty. We developed approaches, in which as an answer returned by the model is not a single value, but the whole distribution of the predicted property. It can facilitate the interpretation of the results and the predictions characterized by high rate of uncertainty are automatically detected. It constitutes great support for medicinal chemists, as it enable fast rejection of compounds with the lowest potential of desired physicochemical/ADMET characteristic.

In addition, for each of the investigated predictive problems, we detected crucial structural features, which should be carefully considered when optimizing compounds towards particular property.

\section{Datasets and fingerprints}
All compounds used in the study were fetched from the ChEMBL database (only data obtained in the human-based experiments were considered).
Then, the compounds were transformed into the bit-string form, using the substructural representation developed by Klekota\&Roth~\cite{klek} (further referred to as KlekFP), consisting of 4860 structural keys.

\begin{figure}[t!]
    \centering
        \includegraphics[width=85mm]{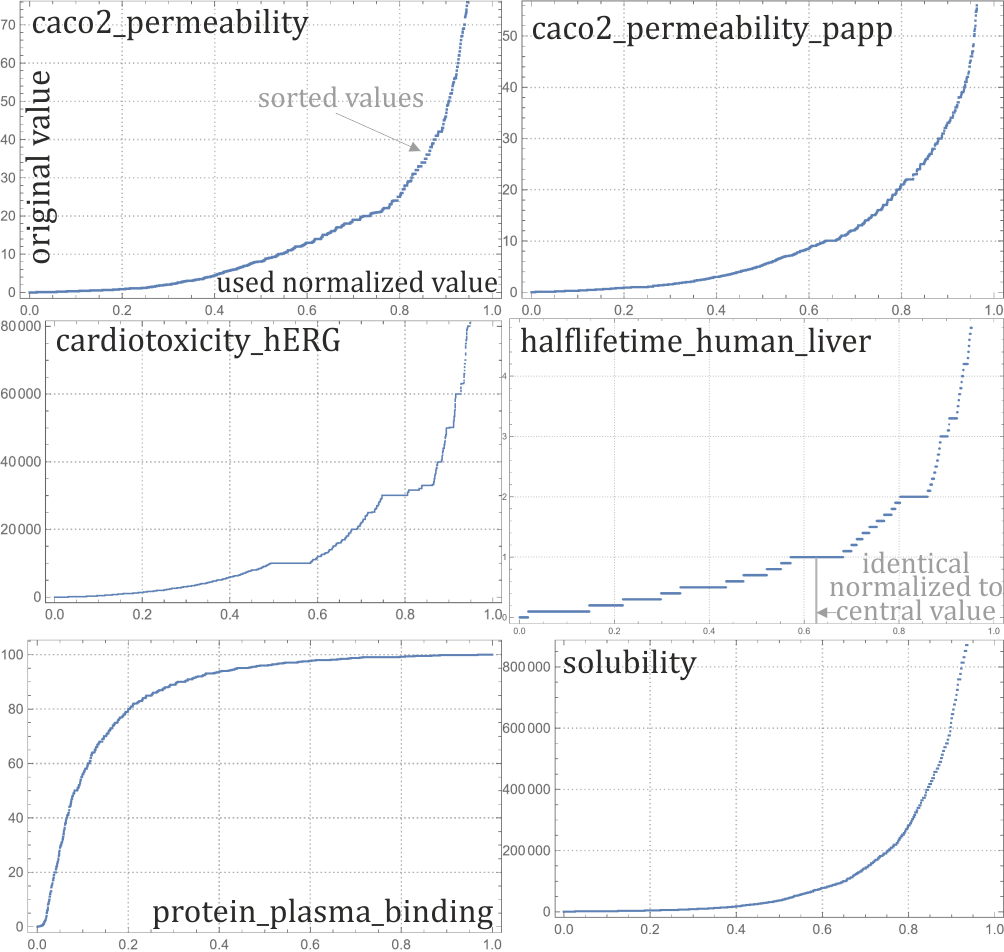}
        \caption{As in copula theory~\cite{copula}, we work on variables normalized  to nearly uniform distribution in $[0,1]$ (close to quantiles) - here using empirical distribution: datapoints are sorted, and each value is assigned its $[0,1]$ position in such order, as in above plots allowing to translate back to the original variables. We can see discreteness: repeating identical values, in which case the normalization assigns the central position of range of identical sorted values.}
       \label{norm}
\end{figure}

\section{Used HCR-based methodology}
This section discusses the used methodology - first normalization here of predicted property to nearly uniform distribution in $[0,1]$, then modeling of its (conditional) probability density as a linear combination in orthonormal basis, with independently predicted coefficients. Regularization techniques are crucial and difficult here - proposed for discussed application, optimized for log-likelihood in 10-fold cross-validation.

\subsection{Normalization}
As in copula theory~\cite{copula}, in the discussed methodology it is convenient to predict density for variable from nearly uniform distribution in $[0,1]$, hence we start with normalization.

Assuming the original variable is $\{z^n\}_{n=1..N}$, such normalization can be done as $y^n =\textrm{CDF}(z^n)$ for CDF being cumulative distribution function, e.g. using some parametric distribution with estimated parameters.

Here the distributions are often quite complex, hence we use empirical distribution instead. Specifically, all $N$ values are first sorted, then $n$-th value in such order is assigned $(n-0.5)/N$ normalized value. If there are multiple identical values, then they are all assigned the central position: $(n_{max}+n_{min}-1)/2N$ - allowing to also work with discrete values.

Normalization for the discussed datasets is presented in Fig. \ref{norm}, also allowing to translate predictions to the original variables. We can see flat ranges corresponding to discretness, which could be included in optimization of the used basis, maybe combined with Canonincal Corralation Analysis as discussed in \cite{hcr3}. Not to complicate it is omitted in this version of article.

\subsection{HCR probability distribution prediction}
In HCR methodology we want to model densities of normalized variables as linear combinations in orthonormal basis, with coefficients ($a_i$) similar to moments: expected value, variance, skewness, kurtosis and higher:
\be\tilde{\rho}(Y=y|X=x)=1+\sum_{i=1}^m f_i(y) a_i(x) \label{e1}\ee
where $y$ is normalized predicted variable, $x=(x_k)_{k=1..K}$ is feature vector - here binary fingerprints (KlekFP, $K=4860$). The $(f_i)_{i=1..m}$ is orthonormal basis: $\int_0^1 f_i(y) f_j(y) dy=\delta_{ij}$, of first $m$ orthonormal polynomials (here $m=20$), starting with ($f_1,f_2,f_3,f_4$):
$$\sqrt{3}(2y-1), \sqrt{5}(6y^2-6y+1), \sqrt{7}(20y^3-30y^2+12y-1),$$
$$3(70y^4-140y^3+90y^2-20y+1). $$

However, $\tilde{\rho}(Y=y|X=x)$ as a linear combination can get below 0. For some applications this issue might be neglected, e.g. in discussed maximal peak analysis trying to find high probability range.

If actual probability density is required, non-negativity is ensured by further calibration: the final density is $\rho(y)=\phi(\tilde{rho}(y))/Z$ where $Z$ is for normalization to integrate to 1, and $phi$ calibration function e.g. $\phi(\rho)=\max(0.1,\rho)$, or some more sophisticated - here there was used parameterized softplus~\cite{softplus} function: \be \phi(\rho)=\ln(1+\exp(\mu x)/\nu)/\mu \ee with $\mu,\nu$ parameters optimized for each dataset. Instead of integration, density was discretized to size 1000 lattice, allowing to use sums instead.

\begin{figure}[t!]
    \centering
        \includegraphics[width=85mm]{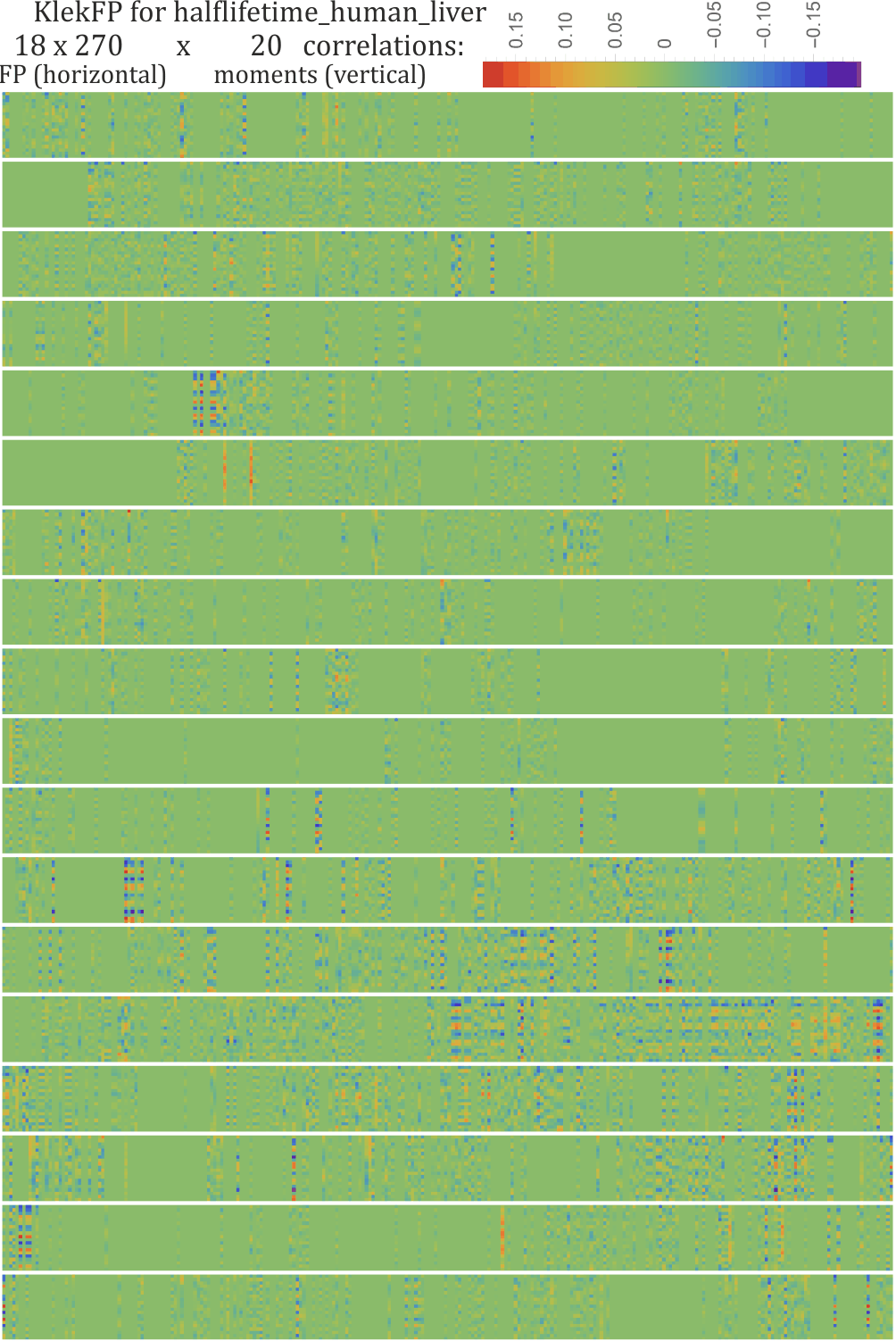}
        \caption{Correlations between $4860=18\cdot 270$ KlekFP fingerprints (horizontal) and $(f_i(y))_{i=1..20}$ moment-like predicted features (vertical) for halflifetime\_human\_liver dataset. We can see many fingerints can be discarded as uncorrelated (green uniform regions) - suggesting feature selection (here with absolute value of correlation above some threshold 0.02-0.04), they can be skipped in calculations to reduce computational costs. In contrast, valuable fingerprints often contribute to multiple moments - providing a deeper understanding of contributions of structural features they describe, also e.g. allowing to improve used $(f_i)$ basis for example with Canonical Correlation Analysis as in \cite{hcr3}.}
       \label{cors}
\end{figure}

Here, as in \cite{hcr1,hcr2,hcr3}, we will directly predict these $a_i$ coefficients with linear regression from feature vector $x$:
\be a_i(x) =\sum_{k=1}^K \beta_{ik} x_k \label{est}\ee
Basic estimation of $a_i$ from $y$ is mean: $\frac{1}{N}\sum_{n=1}^N f_i(y^n)$~\cite{hcr0}. Mean is value minimizing mean squared error from datapoints, hence wanting to predict $a_i$ from variables $x$ here, a natural approach is  mean squared linear regression, in practice with required regularization to prevent overfitting - discussed further.

\subsection{Cross-validation loglik evaluation and regularization}
Discussed model as linear regression is $(\beta_{ik})_{i=1..m, k=1..K}$ coefficients, and its size can easily exceed the number of datapoints $N$, especially if using KlekFP of $K=4860$ fingerprints - hence there are necessary regularization techniques.

From one side it is crucial to use cross-validation evaluation, here $10$-fold cross validation: dataset is randomly split into 10 subsets, 9 of them are treated as training set, 10th as test set - it is done in all 10 ways, and finally averaged.

As evaluation there was directly used log-likelihood: $\frac{1}{N}\sum_{n=1}^N \ln(\rho(y^x|x^n))$. A more informative evaluation is just sorting the $(\rho(y^x|x^n))$ values - presented in Fig. \ref{evs}. There could be also considered more sophisticated evaluations, e.g. directly of performance for a specific application. \\

One regularization technique used in tests is feature selection, here applied separately for each $(a_i)_{i=1..20}$. For this purpose, for each $i=1,\ldots,m$ and $k=1,\ldots,K$ there was calculated correlation between $(f_i(y^n))_{n=1..N}$ predicted vector and $(x^n_k)_{n=1..N}$. Such $K\times m$ correlations are presented in Fig. \ref{cors}. A natural feature selection is choosing those with absolute value of correlation above some threshold ($\approx 0.02-0.04$ ), there is a difficulty of choosing this threshold to remove noise.

For two independent $N$-dimensional vectors, their correlation is approximately from Gaussian distribution centered in zero, and of standard deviation $\sigma=1/\sqrt{N}$. For each $i=1,\ldots,m$ we see $K$ such correlations - we would like to select those of absolute value above some boundary $\xi$. With Gaussian distribution assumption, further than this boundary there should be $\approx p_\xi = 1-\textrm{erf}(\xi/\sigma/\sqrt{2})$ of cases, statistically should happen $\chi \approx K p_\xi$ times. The idea is to fix this $\chi$ (as expect number if uncorrelated) and select features with correlation above boundary $\xi$:
\be \textrm{select}\ |\textrm{correlation}|>\xi\quad \textrm{for}\quad \xi=\sqrt{\frac{2}{N}} \textrm{erf}^{-1}\left(1-\frac{\chi}{K}\right) \label{xi} \ee
To systematize between datasets, there was used $\chi=300$ providing a good compromise. There could be also used e.g. individual optimizations - for datasetes, but also for each $a_i$.

The best performance was obtained by combining above feature selection, with further linear regression with l1 "lasso" regularization:
\be  \forall_{i=1..m}\ \arg\max_{\beta}\  \sum_n \left\|f_i(y^n) - \sum_{k=1}^K \beta_{ik} x_k^n \right\|^2 +\lambda \sum_{k=1}^K |\beta_{ik}| \label{l1}\ee
There is a difficulty to choose the $\lambda$ regularization parameter - here it was separately optimized for each dataset getting $\lambda \approx 3$, it could be optimized separately for each predicted $a_i$, maybe automatically e.g. depending on its number of selected features.
\subsection{Final approach}
\begin{figure}[t!]
    \centering
        \includegraphics[width=85mm]{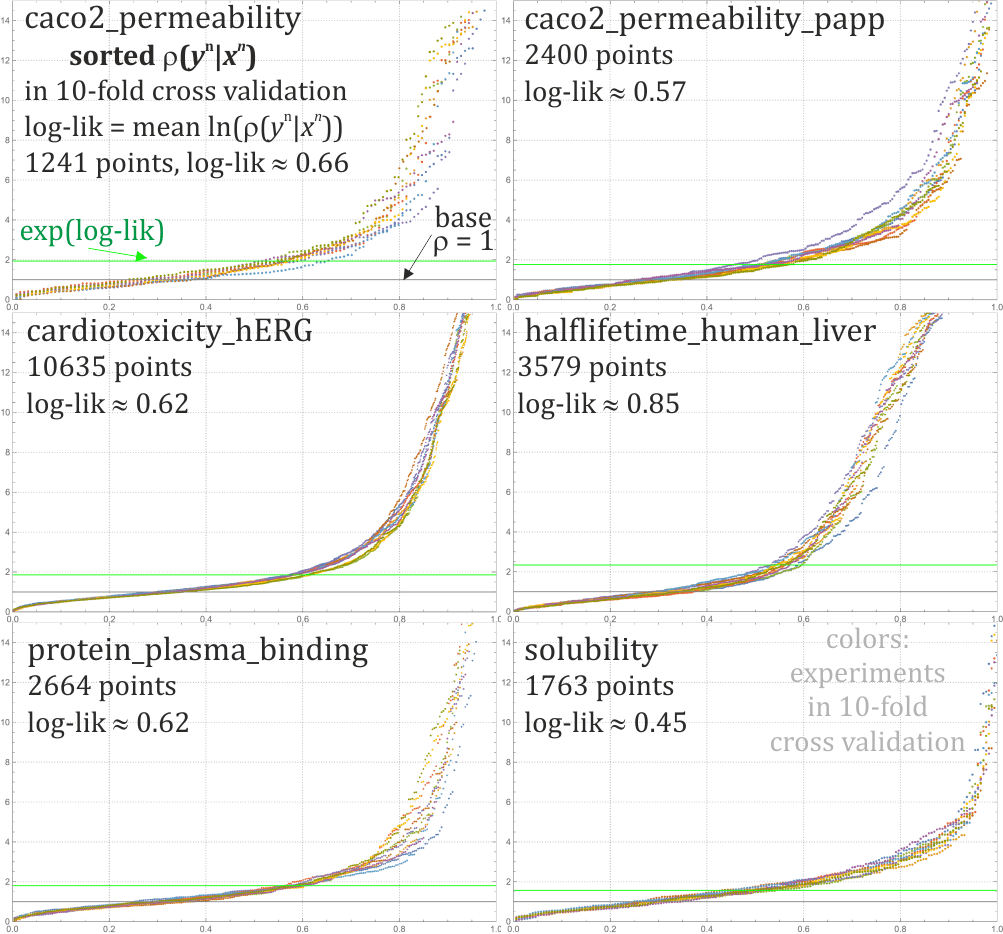}
        \caption{Evaluation of the 6 considered datasets as sorted $\rho(y^n|x^n)$, colors represent 10 divisions into training-test sets in 10-fold cross-validation. We can see that in $\approx 30\%$ of cases it is worse than $\rho=1$ no prediction, what is unavoidable as low probability values also happen sometimes. More importantly, sometimes we have very good localization, e.g. a single peak and real value $y^n$ is indeed in this peak - what seems very useful for virtual screening applications, allowing to restrict to molecules with nearly certain desired parameters. Model parameters were chosen to maximize log-likelihood: $\frac{1}{N}\sum_{n=1}^N \ln(\rho(y^n|x^n))$, however, for a specific application we could directly optimize for its performance. }
       \label{evs}
\end{figure}

Here is summarized the used approach:
\begin{enumerate}
  \item \textbf{Normalize} predicted $(z^n)_{n=1..N}$ variable to $(y^n)_{n=1..N}$ having nearly uniform distribution on $[0,1]$,
  \item For $i=1,\ldots,m$ and $k=1,\ldots,K$ calculate correlation between $(f_i(y^n))_{n=1..N}$ and $(x^n_k)_{n=1..N}$ vectors, \textbf{select fingerprints} with absolute value of correlation above threshold $\xi$ (\ref{xi}),
  \item Among the selected fingerprints, perform \textbf{linear regression with l1 "lasso" regularization} (\ref{l1}),
  \item Having predicted all moments $(a_i)_{i=1..m}$, we \textbf{calculate predicted density} $\tilde{\rho}(y)=1+\sum_{i=1}^m a_i f_i(y)$,
  \item If we need the actual density, finally perform  \textbf{calibration} to enforce non-negativity: $\rho=\phi(\tilde{\rho})/Z$, where $Z=\int_0^1 \phi(\tilde{\rho}(y))dy$ can be calculated on lattice (size 1000 here), and e.g. $\phi(\rho)=\ln(1+\exp(\mu x)/\nu)/\mu $ with $\mu,\nu$ parameters optimized to maximize log-likelihood,
  \item If needed, inverse normalization to translate the predicted density to the original variable (skipped here).
\end{enumerate}
Figure \ref{evs} shows evaluations of such procedure for considered datasets: not only averaged as log-likelihood, but also sorted $\rho(y^n|x^n)$ providing better understanding of predictions: that in $\approx 30\%$ of cases we are worse than no prediction $\rho=1$, but the remaining are usually well localized.

\section{Some possible applications}
This section contains basic discussion of possible applications of predicted probability distribution, for example estimating probability that given property is in the required threshold.

A basic direction is early virtual screening: there is automatically generated a huge number of molecules, and at low computational cost, for further analysis we would like to select the looking more promising ones.

\subsection{Quantile thresholding} 
\begin{figure}[t!]
    \centering
        \includegraphics[width=85mm]{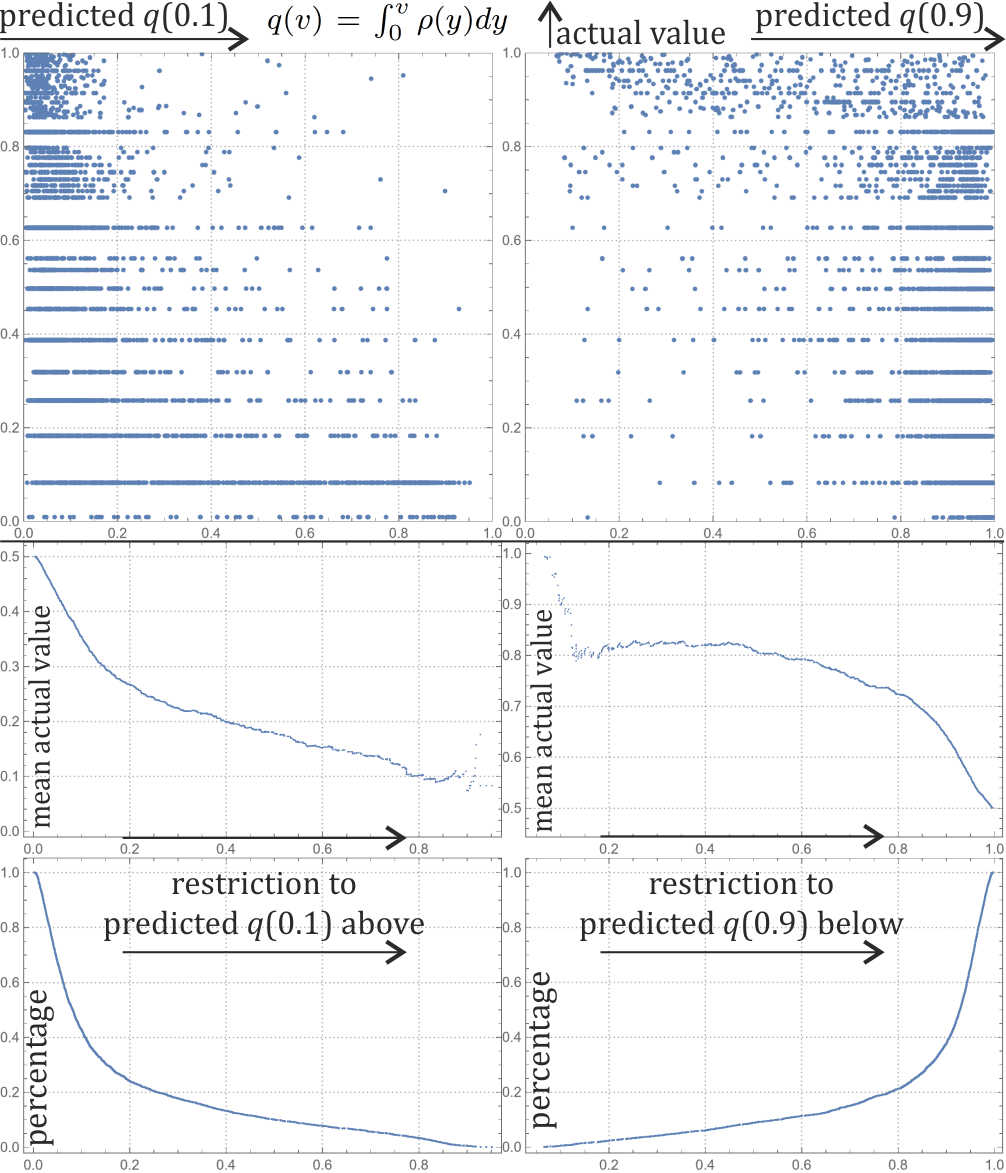}
        \caption{Quantile thresholding: of $q(v)=\int_0^v \rho(y)dy$ for $v=0.1$ (left) and $v=0.9$ (right) for predicted density $\rho$. Top: predicted $q(v)$ (horizontal) vs real value (vertical) for halflifetime\_human\_liver dataset - as expected, we can see that very high values usually have very low predicted $q(0.1)$, very low values have usually very high $q(0.9)$. More practically, predicting very high $q(0.1)$ (or low $q(0.9)$), the real values are nearly always low (high). Plots below allow to evaluate it quantitatively: focusing only on predicted $q(0.1)>0.8$ (horizontal axis), the expected value (horizontal axis) is e.g. $\sim 0.1$. Focusing on predicted $q(0.9)<0.6$, the expected value is e.g. $\sim 0.8$. Plots on the bottom show remaining percentages of cases if performing such restriction. In practice such $v$ choice should be made based on demanded property e.g. minimal/maximal halftime in virtual screening.}
       \label{quantile}
\end{figure}

\begin{figure}[t!]
    \centering
        \includegraphics[width=85mm]{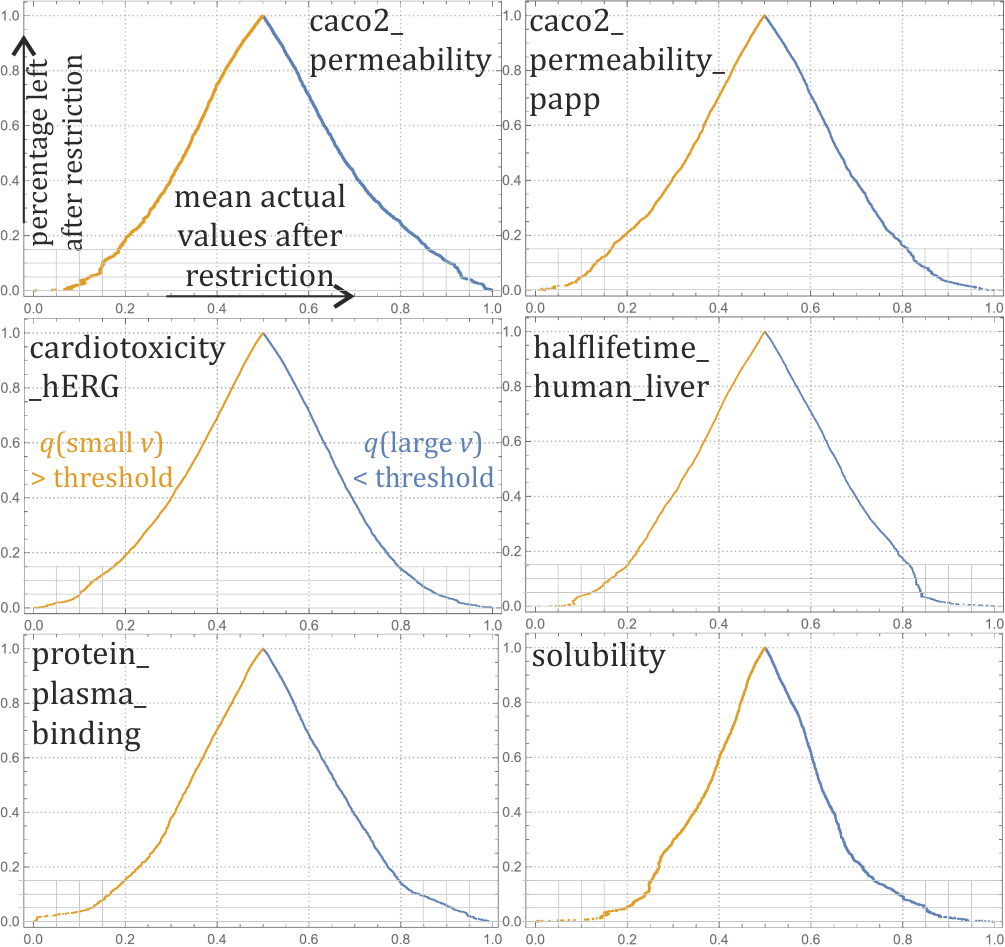}
        \caption{Evaluation of quantile thresholding for all datasets and optimized ($v$, threshold) parameters. Choosing $v$ and restricting to $q(v)=\int_0^v \rho(y)dy$ below (blue) or above (orange) some optimized threshold, the mean actual value was as shown in horizontal axis, at cost of restricting to a percentage of cases shown in vertical axis. For example restriction to $20\%$ e.g. of molecules, allowed to enforce mean value in the highest or lowest $20-30\%$. Restriction to 5\% enforced mean value to extremal 5-15\%. }
       \label{quant}
\end{figure}

Often the basic criterion are molecular properties above or below some threshold. Predicted probability density allows to calculate estimated positions of quantiles:
 \be q(v) = \int_0^v \rho(y) dy \ee
As we could expect and can see in Fig. \ref{quantile}, large $q$ for low $v$ means that the actual value is most likely low, low $q$ for high $v$ means the actual value is most likely high. As in this Figure we can do it quantitatively: choose $v$ and threshold for acceptance of $q(v)$ to get a chosen statistics for the selected ones. Optimizing among $v$ and threshold, Fig. \ref{quant} shows evaluation of such quantile thresholding for all datasets: restring to 20\%, mean of actual value can reach top/bottom 20-30\%. Restring to 5\%, mean of actual values can reach 5-15\%.

While the above is appropriate for filtering objects below/above some threshold, we can also use it to restrict to inside some range: by choosing two bounds $u<v$ and restricting to objects having $q(v)-q(u)$ above some threshold.
\subsection{Maximal peak analysis}
As in Fig. \ref{intr},  especially for high polynomial degree, predicted densities here are often narrow peaks, allowing e.g. to focus on objects (molecules) with this kind of predictions. Here for each predicted density there was found position of its maximum, from it we search left and right until density stops decreasing (or reaching 0,1) - getting left and right local minimum: in $0\leq m_l, m_r\leq 1$. The maximal peak can be defined as $[m_l, m_r]$ range: of width $m_r-m_l$, and predicted probability $\int_{m_l}^{m_r} \rho(y) dy$.

This Figure contains points of (width, probability) of the maximal peaks, marking with color if the real value $y^n$ was indeed in $[m_l, m_r]$. We can see that restricting to objects above some thresholds for width and probability, with very high probability such predicted maximal peak indeed contains the real value - such filtering with thresholds would give a good control of the real value.

\section{Conclusions and further work}
While standard approach is prediction of value, separately predicting multiple moments, here we predict the entire probability distributions - quite successfully for molecular properties from fingerpreints. It allows to inexpensively extract more information, which seems valuable especially for virtual screening applications, e.g. allowing to select a percentage of them with near certainty of chosen given property being in the desired range.\\

This is initial article opening such look new and promising approaches, there are many directions for further development to be explored in the future, for example:
\begin{itemize}
  \item applications in real scenarios especially virtual screening, search for new applications,
  \item improvements of regularization techniques, both from evaluation perspective and computational cost, considering different techniques,
  \item feature engineering - e.g. testing other fingerprints, combining them, calculating only the valuable ones,
  \item building new features, e.g. from fingerprints: $x_{ij}=[x_i = x_j]$ features (1 iff $x_i = x_j$, 0 otherwise), designing completely new ones - also of different types like molecular shape descriptors,
  \item optimizing for different evaluations, e.g. directly of performance for a specific application,
  \item optimizing the $(f_i)$ basis e.g. with Canonical Correlation Analysis, including discreteness - like in \cite{hcr3},
  \item testing more sophisticated prediction techniques for moments, like neural networks.
\end{itemize}
There could be also considered opposite application - Fig. \ref{cors} shows complex dependencies between moments and fingerprints corresponding to structural features, which might be used for their better understanding, or to directly generate molecules of desired properties.

\section{Acknowledgments}
The work of S.P. was supoported by the grant OPUS 2018/31/B/NZ2/00165 financed by the National Science Centre, Poland (www.ncn.gov.pl)

\bibliographystyle{IEEEtran}
\bibliography{cites}
\end{document}